\documentclass[twocolappendix,appendixfloats]{emulateapj}

\usepackage[colorlinks = true, linkcolor = blue, urlcolor  = blue, citecolor = blue, anchorcolor = blue]{hyperref}
\usepackage{hyperref}
\usepackage{apjfonts}
\usepackage{rotating}
\usepackage{amsmath}
\usepackage{color}
\usepackage{color}
\usepackage{CJK}
\usepackage{array}

\newcommand{\pivec}{\mbox{\boldmath $\pi$}}
\newcommand{\muvec}{\mbox{\boldmath $\mu$}}

\newcommand{\thetae}{\theta_{\rm E}}
\newcommand{\pie}{\pi_{\rm E}}
\newcommand{\pien}{\pi_{{\rm E},N}}
\newcommand{\piee}{\pi_{{\rm E},E}}

\definecolor{darkbrown}{RGB}{139,69,19}

\shorttitle{OGLE-2017-BLG-0039}
\shortauthors{Han et al.}

\begin{document}

\title{OGLE-2017-BLG-0039: Microlensing Event 
with Light from the Lens Identified from Mass Measurement}

\author{
C.~Han\altaffilmark{01},
Y.~K.~Jung\altaffilmark{02,201},  
A.~Udalski\altaffilmark{103,202},
I.~Bond\altaffilmark{401,203},
V.~Bozza\altaffilmark{305,306}\\
and\\
M.~D.~Albrow\altaffilmark{04}, S.-J.~Chung\altaffilmark{02,05}, A.~Gould\altaffilmark{02,06,07}, 
K.-H.~Hwang\altaffilmark{02}, D.~Kim\altaffilmark{01}, C.-U.~Lee\altaffilmark{02}, 
H.-W.~Kim\altaffilmark{02}, Y.-H.~Ryu\altaffilmark{02}, I.-G.~Shin\altaffilmark{03}, 
J.~C.~Yee\altaffilmark{03}, Y.~Shvartzvald\altaffilmark{11,201,207},
S.-M.~Cha\altaffilmark{02,10}, S.-L.~Kim\altaffilmark{02,05}, D.-J.~Kim\altaffilmark{02}, 
D.-J.~Lee\altaffilmark{02}, Y.~Lee\altaffilmark{02,10}, B.-G.~Park\altaffilmark{02,05}, 
R.~W.~Pogge\altaffilmark{06}  \\ 
(The KMTNet Collaboration),\\
M.~K.~Szyma\'nski\altaffilmark{103}, P.~Mr\'oz\altaffilmark{103}, J.~Skowron\altaffilmark{103}, 
R.~Poleski\altaffilmark{06,103}, I.~Soszy\'nski\altaffilmark{103}, S.~Koz{\l}owski\altaffilmark{103}, 
P.~Pietrukowicz\altaffilmark{103}, K.~Ulaczyk\altaffilmark{103,104}, M.~Pawlak\altaffilmark{103,105}\\
The OGLE Collaboration),\\
F.~Abe\altaffilmark{409}, R.~Barry\altaffilmark{410}, D.~P.~Bennett\altaffilmark{410,412}, 
A.~Bhattacharya\altaffilmark{410,412}, M.~Donachie\altaffilmark{413}, P.~Evans\altaffilmark{413}, 
A.~Fukui\altaffilmark{414}, Y.~Hirao\altaffilmark{422}, Y.~Itow\altaffilmark{409}, 
K.~Kawasaki\altaffilmark{422}, N.~Koshimoto\altaffilmark{422}, M.~C.~A.~Li\altaffilmark{413}, 
C.~H.~Ling\altaffilmark{401}, Y.~Matsubara\altaffilmark{409}, S.~Miyazaki\altaffilmark{422}, 
H.~Munakata\altaffilmark{409}, Y.~Muraki\altaffilmark{409}, M.~Nagakane\altaffilmark{422}, 
K.~Ohnishi\altaffilmark{415}, C.~Ranc\altaffilmark{410}, N.~Rattenbury\altaffilmark{413}, 
T.~Saito\altaffilmark{416}, A.~Sharan\altaffilmark{413}, D.~J.~Sullivan\altaffilmark{415}, 
T.~Sumi\altaffilmark{422}, D.~Suzuki\altaffilmark{418}, P.~J.~Tristram\altaffilmark{419}, 
T.~Yamada\altaffilmark{422}, A.~Yonehara\altaffilmark{422}\\
(The MOA Collaboration)\\
}

\email{cheongho@astroph.chungbuk.ac.kr}

\altaffiltext{01}{Department of Physics, Chungbuk National University, Cheongju 28644, Republic of Korea} 
\altaffiltext{02}{Korea Astronomy and Space Science Institute, Daejon 34055, Republic of Korea}
\altaffiltext{03}{Harvard-Smithsonian Center for Astrophysics, 60 Garden St., Cambridge, MA, 02138, USA}   
\altaffiltext{04}{University of Canterbury, Department of Physics and Astronomy, Private Bag 4800, Christchurch 8020, New Zealand} 
\altaffiltext{05}{Korea University of Science and Technology, 217 Gajeong-ro, Yuseong-gu, Daejeon 34113, Republic of Korea} 
\altaffiltext{06}{Department of Astronomy, Ohio State University, 140 W.\ 18th Ave., Columbus, OH 43210, USA} 
\altaffiltext{07}{Max Planck Institute for Astronomy, K\"onigstuhl 17, D-69117 Heidelberg, Germany} 
\altaffiltext{10}{School of Space Research, Kyung Hee University, Yongin 17104, Republic of Korea} 
\altaffiltext{11}{Jet Propulsion Laboratory, California Institute of Technology, 4800 Oak Grove Drive, Pasadena, CA 91109, USA} 
\altaffiltext{305}{Dipartimento di Fisica ``E.~R.~Caianiello'', Universitá di Salerno, Via Giovanni Paolo II, I-84084 Fisciano (SA) , Italy} 
\altaffiltext{306}{Istituto Nazionale di Fisica Nucleare, Sezione di Napoli, Via Cintia, I-80126 Napoli, Italy} 
\altaffiltext{103}{Warsaw University Observatory, Al.~Ujazdowskie 4, 00-478 Warszawa, Poland} 
\altaffiltext{104}{Department of Physics, University of Warwick, Gibbet Hill Road, Coventry, CV4 7AL, UK}
\altaffiltext{105}{Institute of Theoretical Physics, Faculty of Mathematics and Physics, Charles University in Prague, Czech Republic}
\altaffiltext{401}{Institute of Natural and Mathematical Sciences, Massey University, Auckland 0745, New Zealand} 
\altaffiltext{409}{Institute for Space-Earth Environmental Research, Nagoya University, 464-8601 Nagoya, Japan} 
\altaffiltext{410}{Code 667, NASA Goddard Space Flight Center, Greenbelt, MD 20771, USA} 
\altaffiltext{412}{Deptartment of Physics, University of Notre Dame, 225 Nieuwland Science Hall, Notre Dame, IN 46556, USA} 
\altaffiltext{413}{Department of Physics, University of Auckland, Private Bag 92019, Auckland, New Zealand} 
\altaffiltext{414}{Okayama Astrophysical Observatory, National Astronomical Observatory of Japan, Asakuchi, 719-0232 Okayama, Japan} 
\altaffiltext{415}{Nagano National College of Technology, 381-8550 Nagano, Japan} 
\altaffiltext{416}{Tokyo Metroplitan College of Industrial Technology, 116-8523 Tokyo, Japan} 
\altaffiltext{417}{School of Chemical and Physical Sciences, Victoria University, Wellington, New Zealand} 
\altaffiltext{418}{Institute of Space and Astronautical Science, Japan Aerospace Exploration Agency, Kanagawa 252-5210, Japan} 
\altaffiltext{419}{Mt.~John University Observatory, P.O. Box 56, Lake Tekapo 8770, New Zealand} 
\altaffiltext{422}{Department of Earth and Space Science, Graduate School of Science, Osaka University, Toyonaka, Osaka 560-0043, Japan} 
\altaffiltext{201}{The KMTNet Collaboration} 
\altaffiltext{202}{The OGLE Collaboration} 
\altaffiltext{203}{The MOA Collaboration} 
\altaffiltext{207}{NASA Postdoctoral Program Fellow}

\begin{abstract}
We present the analysis of the caustic-crossing binary microlensing event OGLE-2017-BLG-0039.  
Thanks to the very long duration of the event, with an event time scale $t_{\rm E}\sim 130$ 
days, the microlens parallax is precisely measured despite its small value of $\pie\sim 0.06$.  
The analysis of the well-resolved caustic crossings during both the source star's entrance and 
exit of the caustic yields the angular Einstein radius $\thetae\sim 0.6$~mas.  The measured 
$\pie$ and $\thetae$ indicate that the lens is a binary composed of two stars with masses 
$\sim 1.0~M_\odot$ and $\sim 0.15~M_\odot$, and it is located at a distance of $\sim 6$ kpc.
From the color and brightness of the lens estimated from the determined lens mass and distance, 
it is expected that $\sim 2/3$ of the $I$-band blended flux comes from the lens.  Therefore, 
the event is a rare case of a bright lens event for which 
high-resolution follow-up observations 
can confirm the nature of the lens.
\end{abstract}

\keywords{gravitational lensing: micro -- binaries: general}

\section{Introduction}\label{section:one}

It is believed that most microlensing events detected toward the Galactic bulge field are produced 
by stars \citep{Han2003b}. For stellar lens events, the observed light comes from the lens as well 
as from the source star. However, lenses, in most cases, are much fainter than the source stars being 
monitored, and thus it is difficult to detect the light from the lens. In some rare cases for which 
the lenses are bright, the light from the lens can be detected. However, even in such cases, it is 
still difficult to attribute the excess flux to the lens because the flux may come from nearby
stars blended in the image of the source or possibly from the companion to the 
source if the source is a binary.

\begin{figure*}
\epsscale{0.81}
\plotone{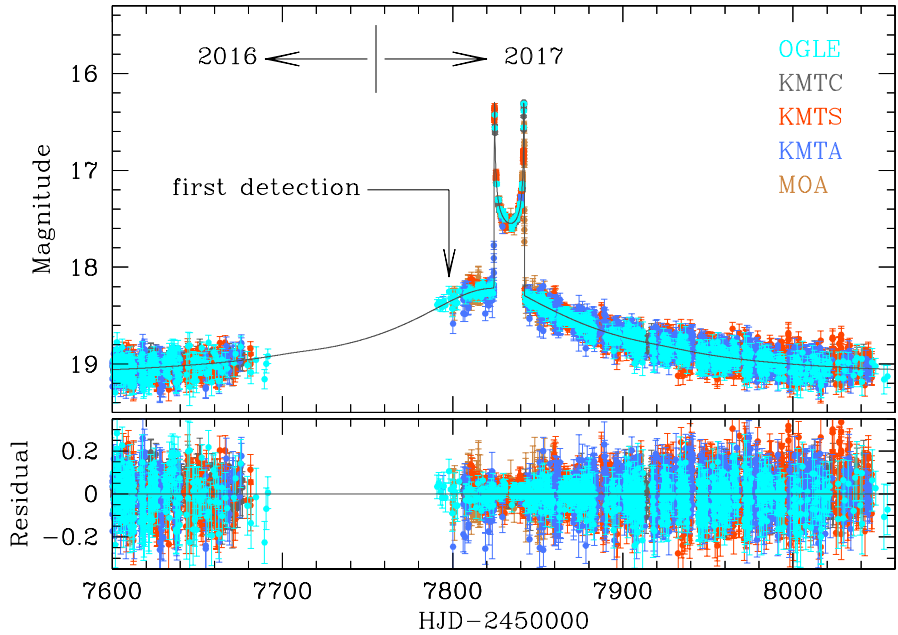}
\caption{
Light curve of OGLE-2017-BLG-0039. The curve superposed on data points represents the best-fit 
binary lens model (close binary solution with $u_0>0$). The lower panel shows the residual from 
the model.
}
\label{fig:one}
\end{figure*}

There have been several methods proposed to identify light from lenses. The most explicit method is 
resolving the lens from the source or other blended stars from high-resolution follow-up observations. 
For typical lensing events, the relative lens-source proper motions is $\mu\sim 5$ mas yr$^{-1}$. 
This means that one has to wait $\sim 10$ -- 20 years for direct lens imaging until the lens is 
separated enough from the source even using the currently available instrument with the highest 
resolution \citep{Han2003a}.\footnote{With adaptive optics observations using the Extremely Large 
Telescope, which is planned to operate in 2024, the waiting time will be reduced into several years, 
and direct lens imaging will become routine.} As a result, the method has been applied to only 3 
lensing events: MACHO-LMC-5 \citep{Alcock2001}, MACHO-95-BLG-37 \citep{Kozlowski2007}, and 
OGLE-2005-BLG-169 \citep{Batista2015, Bennett2015}.

A bright lens can also be identified from astrometric observations of lensing events. When a source 
star is gravitationally lensed, the centroid of the source star image is displaced from the position 
of the unlensed source star \citep{Hog1995, Miyamoto1995, Paczynski1998, Boden1998}.  Due to the 
relative lens-source motion, the position of the image centroid traces out an elliptical trajectory 
where the size and shape of the trajectory are determined by the angular Einstein radius and the impact 
parameter of the lens-source approach \citep{Walker1995, Jeong1999}. If an event is produced by a bright 
lens, the astrometric shift is affected by the light from the lens and one can identify the bright lens 
from the deviation \citep{Han1999}.    
Lensing-induced astrometric shifts produced by very nearby stars have been measured, e.g., 
Stein 2051 B \citep{Sahu2017} and Proxima Centauri \citep{Zurlo2018}.  For general lensing events, 
however, it is currently difficult to measure image motions due to the required high accuracy of 
an order 10 $\mu$as.
 
In some limited cases, a bright lens can also be identified from the analysis of the lensing light 
curve obtained from photometric observations. This photometric identification is possible for events 
where the mass $M$ and distance $D_{\rm L}$ to the lens are determined.  With the determined $M$ and 
$D_{\rm L}$, one can predict the color and brightness of the lens.  If the color and brightness are 
close to those of the blend, then, it is likely that the flux from the lens comprises an important 
portion of the total blended flux.  Due to the bright nature of the lens, the lens will be visible 
in high-resolution images, which are obtained from space-based or ground-based adaptive optic (AO) 
observations, as an additional light that is blended with the source image \citep{Bennett2007}.  
Then, one can identify the lens by comparing the excess flux with the prediction from lensing modeling.  
Bright lenses have been identified by space-based follow-up observations using {\it Hubble Space Telescope} 
for the first two planetary microlensing events OGLE-2003-BLG-235 \citep{Bond2004} and OGLE-2005-BLG-071 
\citep{Udalski2005} by \citet{Bennett2006} and \citet{Dong2009}, respectively.  For OGLE-2006-BLG-109 
\citep{Gaudi2008, Bennett2010} and OGLE-2012-BLG-0026 \citep{Han2013}, for which multiple microlensing 
planetary systems were found, the light from the lenses were confirmed by Keck AO follow-up observations 
conducted by \citet{Bennett2010} and \citet{Beaulieu2016}, respectively.

For various reasons, long time-scale binary-lens events are ideal targets in applying the photometric 
method of bright lens identification. First, the chance to determine the lens mass and distance is 
high for these events. For the determinations of $M$ and $D_{\rm L}$, it is required to measure both 
the microlens parallax $\pie$ and the angular Einstein radius $\thetae$, which are related to $M$ and 
$D_{\rm L}$ by
\begin{equation}
M={\thetae\over \kappa\pie}
\label{eq1}
\end{equation}
and
\begin{equation}
D_{\rm L}={{\rm au}\over \pie\thetae +\pi_{\rm S}}
\label{eq2}
\end{equation}
respectively \citep{Gould2000}. Here $\kappa=4G/(c^2{\rm au})$, $\pi_{\rm S}={\rm au}/D_{\rm S}$ 
is the parallax of the source star, and $D_{\rm S}$ denotes the distance to the source.
We note that $\pi_{\rm S}$ is known for events detected toward the Galactic bulge direction because 
source stars are located in the bulge and thus $D_{\rm S}$ is known. 
The angular 
Einstein radius is measured from the deviation in lensing light curves affected by finite-source 
effects.  For single-lens events, finite-source effects can be detected only for very rare events in 
which the lens passes over the surface of the source star, e.g., \citet{Choi2012}, and thus the chance 
to measure $\thetae$ is low.\footnote{We note that although rare, source-passing single-lens events 
are important because they provide a channel to measure the masses of single mass objects.  In addition 
to $\thetae$ from finite-source effects, the lens masses in two cases of extremely high-magnification 
events \citep{Gould2009, Yee2009} are determined by ``terrestrial parallax'' measurement of $\pie$ 
\citep{Gould1997}.  The chance to measure $\pie$ is becoming higher as more events are simultaneously 
observed from the ground and from a space-based satellite in a heliocentric orbit such as the 
{\it Spitzer} telescope: space-based microlens parallax \citep{Refsdal1966, Gould1994}.  The events 
with measured lens masses from {\it Spitzer} observations include OGLE-2015-BLG-1268, OGLE-2015-BLG-0763 
\citep{Zhu2016}, OGLE-2017-BLG-0896 \citep{Shvartzvald2018}, OGLE-2017-BLG-1186, OGLE-2017-BLG-0722, 
OGLE-2017-BLG-1161, OGLE-2017-BLG-1254 \citep{Zang2018}, OGLE-2015-BLG-1482 \citep{Chung2017}, 
OGLE-2016-BLG-1045 \citep{Shin2018}.} For binary-lens events, in contrast, the chance to measure 
$\thetae$ is high because these events usually accompany caustic-crossing features in lensing light 
curves from which one can measure $\thetae$. The chance to measure $\pie$ is also high for these events 
due to their long time scales. As an event time scale approaches or exceeds the orbital period of Earth, 
i.e., 1 yr, the relative lens-source motion deviates from rectilinear due to the acceleration of the 
observer's motion caused by the Earth's orbital motion. The deviation in the relative lens-source motion 
results in lensing light curve deviations from which one can measure the microlens parallax \citep{Gould1992}. 
Being able to measure both $\thetae$ and $\pie$, the chance to measure the physical lens parameters of $M$ 
and $D_{\rm L}$ is high for long time-scale binary-lens events. Second, the chance for these events to be 
produced by bright lenses is higher.  This is because the event time scale is proportional to the square 
root of the lens mass, i.e., $t_{\rm E}\propto \sqrt{M}$, and thus the lens mass tends to be heavier and 
brighter than general events with $t_{\rm E}\sim (O)10$ days.  As the lens becomes brighter, the 
fraction of the lens flux in the total blended flux increases, making it easier to identify the bright lens.

In this work, we present the analysis of the binary-lens event OGLE-2017-BLG-0039.  The event has a 
very long time scale with well-resolved caustic-crossing features in the light curve. Being able to 
determine the lens mass and distance by simultaneously  measuring both $\pie$ and $\thetae$, we check 
the lens origin of the blended light.

\section{Observations}\label{section:two}

The microlensing event OGLE-2017-BLG-0039 occurred on a star located toward the Galactic bulge field. The 
coordinates of the source star are $({\rm RA}, {\rm DEC})_{\rm J2000}=$ (18:01:47.95, -27:20:34.7), which 
corresponds to the galactic coordinates $(l, b)= (3.18^\circ, -2.27^\circ)$. The apparent baseline brightness 
before lensing magnification was $I_{\rm base}\sim 19.1$.

The lensing-induced brightening of the source star was found and alerted by the Early Warning System 
of the Optical Gravitational Lensing Experiment \citep[OGLE;][]{Udalski2015} survey on 2017 February 
14 (${\rm HJD}^\prime={\rm HJD}-2450000\sim 7798$). The OGLE survey was conducted using the 1.3 m 
Warsaw Telescope located at the Las Campanas Observatory in Chile. The event was in the OGLE BLG511.23 
field, toward which observations were carried out 3 -- 10 times per night. Images were taken primarily 
in $I$ band and $V$-band images were occasionally obtained for the color measurement of the source star. 
Photometry of the images is extracted using the OGLE difference-imaging pipeline \citep{Udalski2003}.

The event was also in the fields that were observed by the Microlensing Observations in Astrophysics 
\citep[MOA:][]{Bond2001} and the Korea Microlensing Telescope Network \citep[KMTNet:][]{Kim2016} 
surveys. MOA observations were conducted in a customized $R$ band using the 1.8~m telescope of the 
Mt.~John University Observatory in New Zealand. The event was dubbed MOA-2017-BLG-080 in the list of 
MOA transient events. About 10 images were taken each night from the MOA survey. KMTNet observations of 
the event were conducted using three identical 1.6 m telescopes that are globally distributed at the 
Cerro Tololo Interamerican Observatory, Chile (KMTC), the South African Astronomical Observatory, South 
Africa (KMTS), and the Siding Spring Observatory, in Australia (KMTA). Most KMTNet data were taken in 
$I$ band, and some $V$-band images were acquired for the source color measurement.  The event is in the 
KMTNet BLG03 field for which observations were conducted at a cadence of 2 per hour.  Most of the BLG03 
field is additionally covered by the BLG43 field with a slight offset in order to cover the gaps between 
CCD chips as well as to increase the rate of observation.  The event happens to be just outside
the BLG43 field, and thus no data is obtained from the BLG43 field.  Photometry of the MOA and KMTNet 
data was conducted using the software packages customized by the individual groups based on the difference 
imaging method: \citet{Bond2001} and \citet{Albrow2017} for the MOA and KMTNet surveys, respectively.

Figure~\ref{fig:one} shows the light curve of OGLE-2017-BLG-0039. It shows that the lensing magnification 
was in progress from the second half of the 2016 season and lasted throughout the 2017 season.  The event could 
not be observed during roughly three-month period when the Sun passed the bulge field.  The event was found in 
the early 2017 season when the source was brightened by $\sim 0.7$ mag.  On 2017 March 12, the source became 
brighter by $\sim 2$ mag within $\lesssim  0.5$ day. Such a sudden brightening occurs when a source passes 
over the caustic formed by a binary lens.  The binary-lens interpretation became more plausible as the 
light curve exhibited a ``U''-shape feature after the caustic-crossing spike.  Because caustics of a binary 
lens form closed curves, caustic-crossings occur in pairs.  On 2017 March 21 (${\rm HJD}^\prime\sim 7833$), 
V.~Bozza circulated a model based on the binary-lens interpretation and predicted that another caustic 
crossing would occur on ${\rm HJD}^\prime\sim 7842$.  The second caustic crossing occurred at about the 
anticipated time. Bozza circulated another model based on additional data after the second caustic crossing. 
After completing the caustic crossings, the event gradually returned to its baseline.

\begin{figure}
\includegraphics[width=\columnwidth]{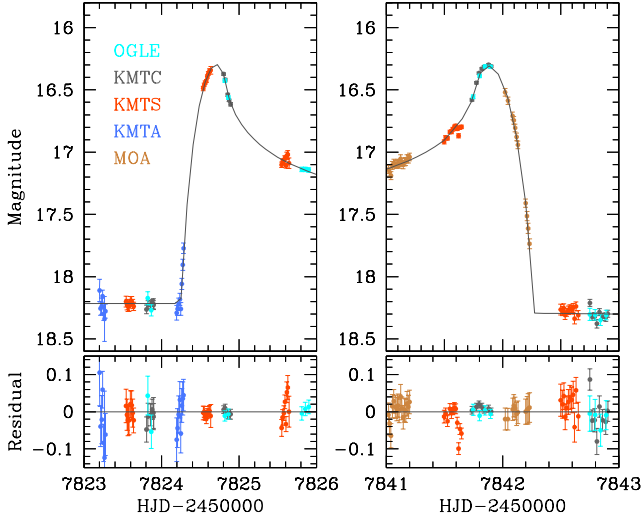}
\caption{
Zoom of lensing light curve around the caustic crossing regions.  The left and right panels show the 
light curve when the source entered and exited the caustic, respectively.  The bottom panels show 
the residual from the best-fit model.
}
\label{fig:two}
\end{figure}

\begin{table}
\centering
\caption{Data used in the analysis}
\begin{tabular}{lccc}
\hline\hline
\multicolumn{2}{c}{Data set} &
\multicolumn{1}{c}{Coverage (HJD$^\prime$)} &
\multicolumn{1}{c}{$N_{\rm data}$} \\
\hline
OGLE    &        &  5376 -- 8236   &  5751  \\
KMTNet  & KMTA   &  7444 -- 8220   &  2237  \\
        & KMTC   &  7439 -- 8216   &  2842  \\
        & KMTS   &  7441 -- 8216   &  3739  \\
MOA     &        &  7800 -- 7860   &  268   \\
\hline
\end{tabular}
\label{table:one}
\tablecomments{ 
${\rm HJD}^\prime={\rm HJD}-2450000$.
}
\end{table}

Two factors make the event scientifically important. The first factor is that the duration of the event 
is very long. The lensing-induced magnification of the source flux started in the 2016 season and lasted 
throughout the 2017 season. Due to the long duration of the event, the chance to measure the microlens 
parallax is high. The second factor is that both of the caustic crossings when the source entered and 
exited the caustic were densely and continuously covered from the combined observations  
using the globally distributed telescopes. In Figure~\ref{fig:two}, we present the zoom of the light curve 
during the caustic entrance (left panels) and exit (right panels). Analysis of the caustic-crossing parts 
of the light curve enables one to measure the angular Einstein radius. Being able to measure both 
$\pie$ and $\thetae$, then, one can uniquely determine the mass and distance to the lens. 
This also enables one to check the lens origin of the blended flux.

In Table~\ref{table:one}, we list details about the data used in the analysis. The coverage column 
indicates the time range, and $N_{\rm data}$ represents the number of data points in the individual 
data sets. For the OGLE data set, we use 9 years of data from 2010 to 2018 for the secure measurement 
of the baseline magnitude.  Although the event was observed by the KMTNet survey since its commencement 
in 2015, the system was under development during the 2015 season.  We, therefore, use KMTNet data that 
have been acquired since 2016 season after the system was stabilized.  For the MOA data, photometric 
uncertainties of the data near the baseline are considerable, but the data densely covered the caustic 
exit.  We, therefore, use partial MOA data around caustic-crossing features obtained during 
$7800\lesssim {\rm HJD}^\prime\lesssim 7860$.

\section{Analysis}\label{section:three}
 
Considering the caustic-crossing features, we model the observed light curve based on the binary-lens 
interpretation. We start modeling under the assumption that the relative lens-source motion is rectilinear 
although it is expected that the motion would be non-rectilinear due to the long duration of the event. Under 
this assumption, a lensing light curve is described by 7 principal parameters. Four of these parameters 
describe the lens-source approach including the time of the closest source approach to a reference position 
of the lens, $t_0$, the source-reference separation at that time, $u_0$ (impact parameter), the event time 
scale, $t_{\rm E}$, and the angle between the source trajectory and the binary axis, $\alpha$ (source trajectory 
angle).  For the reference position of the lens, we use the center of mass of the binary lens system.  Another 
two parameters describe the binary lens including the projected binary separation normalized to $\thetae$, $s$, 
and the mass ratio between the binary lens components, $q$. The last parameter is the normalized source radius 
$\rho$, which is defined as the ratio of the angular source radius $\theta_*$ to the angular Einstein radius, 
i.e., $\rho=\theta_*/\thetae$.

In the preliminary modeling, we conduct a grid search for the binary-lens parameters $s$ and $q$ while other 
parameters are searched for using a downhill approach. For the downhill approach, we use the Markov Chain 
Monte Carlo (MCMC) method. From this preliminary search, we find two candidate solutions with 
$(s,q)_{\rm close}\sim (0.84, 0.13)$ and $(s,q)_{\rm wide}\sim (1.61, 0.20)$. We designate the individual 
solutions as ``close'' and ``wide'' based on the fact that the projected binary separation is less ($s<1$) 
and greater ($s>1$) than the angular Einstein radius, respectively.

Although the solutions found from the preliminary modeling describe the overall light curve, they 
leave subtle long-term residuals from the models. It is known that long-term deviations are caused 
by two major higher-order effects. The first one is the microlens-parallax effect that is caused by 
the orbital motion of Earth \citep{Gould1992}.  The other one is caused by the orbital motion of the 
lens, lens-orbital effects \citep{Dominik1998, Ioka1999}. We, therefore, check whether the fit improves 
with the consideration of these higher-order effects.

\begin{table}
\centering
\caption{Comparison of models}
\begin{tabular}{lccc}
\hline\hline
\multicolumn{2}{c}{Model} &
\multicolumn{2}{c}{$\chi^2$} \\
\multicolumn{2}{c}{} &
\multicolumn{1}{c}{Close} &
\multicolumn{1}{c}{Wide} \\
\hline
Static          &             &  16081.8  &  15634.0   \\
Orbit           &             &  15476.8  &  15565.7   \\
Parallax        & ($u_0>0$)   &  15886.7  &  15565.6   \\
--              & ($u_0<0$)   &  15862.8  &  15575.4   \\
Orbit+Parallax  & ($u_0>0$)   &  15355.8  &  15479.2   \\
--              & ($u_0<0$)   &  15362.2  &  15463.4   \\
\hline
\end{tabular}
\label{table:two}
\end{table}

Consideration of the microlens-parallax effect requires one to include two additional lensing parameters 
of $\pien$ and $\piee$. They represent the north and east components of the microlens-parallax vector, 
$\pivec_{\rm E}$, projected onto the sky along the north and east equatorial coordinates, respectively. 
The magnitude of the microlens-parallax vector is $\pie=(\pien^2+\piee^2)^{1/2}$ and it is directed 
toward the same direction as that of the relative lens-source proper motion vector $\muvec$, i.e., 
$\pivec_{\rm E}=\pie(\muvec/\mu)$.  Consideration of the lens-orbital effect also requires additional 
parameters. Under the approximation that the change of the lens position is small, the effect is 
described by two parameters of $ds/dt$ and $d\alpha/dt$, which represent the rates of change in the 
projected binary separation and the source trajectory angle, respectively.

\begin{figure}
\includegraphics[width=\columnwidth]{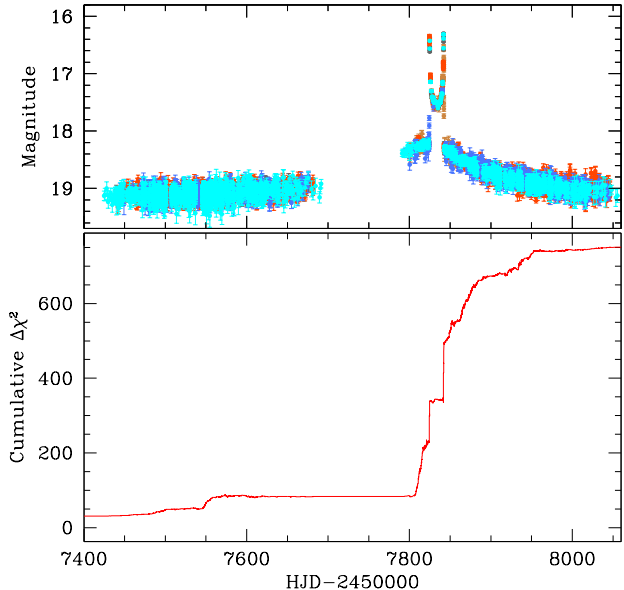}
\caption{
Cumulative distribution of $\Delta\chi^2$ between the models with and without the consideration of 
the microlens-parallax and lens-orbital effects. The light curve in the upper panel is presented 
to show the region of fit improvement.
}
\label{fig:three}
\end{figure}

In Table~\ref{table:two}, we list the results of the additional modeling considering the higher-order 
effects. We compare the goodness of the fit for the individual models in terms of $\chi^2$ values. The 
``static'' model denotes the solution obtained under the assumption of the rectilinear relative lens-source 
motion.  In the ``orbit'' and ``parallax'' models, we separately consider the microlens-parallax and the lens-orbital 
effects, respectively. In the ``orbit+parallax'' model, we simultaneously consider both higher-order effects. 
For microlensing solutions obtained considering microlens-parallax effects, it is known that there may exist 
a pair of degenerate solutions with $u_0>0$ and $u_0<0$ due to the mirror symmetry of the source trajectories
with respect to the binary axis between the two degenerate solutions: ecliptic degeneracy 
\citep{Smith2003, Skowron2011}. We, therefore, check the ecliptic degeneracy whenever parallax effects are 
considered in modeling. The lensing parameters of the pair of solutions caused by the ecliptic degeneracy 
are roughly in the relation 
$(u_0, \alpha, \pien, d\alpha/dt) \leftrightarrow -(u_0, \alpha, \pien, d\alpha/dt)$.

From the comparison of $\chi^2$ values between the close and wide binary solutions, it is found that 
the close solution is preferred over the wide solution by $\Delta\chi^2\sim 123$.  This is statistically 
significant enough to resolve the degeneracy between the two solutions.

We find that consideration of the higher-order effects significantly improves the fit. It is found 
that the higher-order effects improve the fit by $\Delta\chi^2\sim 726$ for the close-binary solution, 
which provides the best-fit solution. For the pair of solutions resulting from the ecliptic degeneracy, 
it is found that the degeneracy is moderately severe with $\Delta\chi^2\sim 6.4$, and the $u_0>0$ 
solution is preferred over the $u_0<0$ solution.  In order to better show the region of fit improvement 
by the higher-order effects, in Figure~\ref{fig:three}, we present the cumulative distribution of 
$\Delta\chi^2$, where $\Delta\chi^2$ represents the difference in $\chi^2$ between the static and 
orbit+parallax ($u_0>0$) models. It shows that the fit improvement occurs throughout the event.

\begin{figure}
\includegraphics[width=\columnwidth]{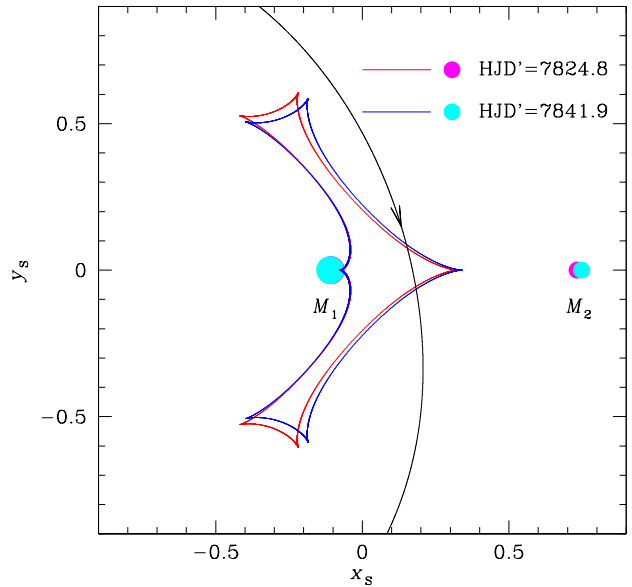}
\caption{
Configuration of the lens system. The curve with an arrow represents the source trajectory. The closed 
curve composed of 6 folds represent the caustic. The small filled dots marked by $M_1$ and $M_2$ represent 
the positions of the binary-lens components.  All lengths are scaled to the angular Einstein radius 
corresponding to the total mass of the lens.  Due to the orbital motion of the lens, the positions of 
the lens components and the caustic shape vary in time. We mark the positions at HJD$^\prime$=7824.8 and 
7841.9, which correspond to the times of source star's caustic entrance and exit, respectively.
}
\label{fig:four}
\end{figure}

In Table~\ref{table:three}, we list the lensing parameters determined from modeling. Although the 
close/wide degeneracy is clearly resolved, we present the parameters of the wide-binary solution for 
readers who may want to reproduce the result. In Figure~\ref{fig:four}, we also present the lens-system 
configuration in which the source trajectory (curve with an arrow) with respect to the binary lens 
components (marked by $M_1$ and $M_2$) and the caustic (cuspy closed figure) is shown. According to 
the best-fit model (close $u_0>0$ solution), it is found that the lensing event was produced by a 
binary with $(s, q)\sim (0.85, 0.15)$.  Due to the proximity of the binary separation to the angular 
Einstein radius, the caustics form a single closed curve composed of 6 folds. Due to the relatively 
low mass ratio, $q\sim 0.15$, the caustic is tilted toward heavier-mass lens component, i.e., $M_1$, 
with a protruding cusp pointing toward the lower-mass lens component, $M_2$. The source passed through 
the protruding cusp region of the caustic, producing the two observed spikes when it entered and exited 
the caustic. The source trajectory is curved by the higher-order effects, especially by the lens-orbital 
effect. The change rate of the source trajectory angle is 
$|d\alpha/dt|\sim 1.22~{\rm rad}~{\rm yr}^{-1}\sim 70^\circ~{\rm yr}^{-1}$.

\begin{table*}
\centering
\caption{Lensing parameters}
\begin{tabular}{m{2.5cm} m{2.5cm} m{2.5cm} m{2.5cm} m{2.5cm} }
\hline\hline
\multicolumn{1}{c}{Parameter} &
\multicolumn{2}{c}{Close} &
\multicolumn{2}{c}{Wide} \\
\multicolumn{1}{c}{} &
\multicolumn{1}{c}{$u_0>0$} &
\multicolumn{1}{c}{$u_0<0$} &
\multicolumn{1}{c}{$u_0>0$} &
\multicolumn{1}{c}{$u_0<0$} \\
\hline
$\chi^2$                       & 15355.8               & 15362.2                & 15479.2                 & 15463.4                \\  
$t_0$ (HJD$^\prime$)           & 7829.104 $\pm$ 0.157  &  7829.248 $\pm$ 0.050  &  7834.893 $\pm$ 0.018   &  7834.209 $\pm$ 0.206  \\
$u_0$                          &    0.167 $\pm$ 0.003  &    -0.161 $\pm$ 0.001  &     0.026 $\pm$ 0.001   &    -0.007 $\pm$ 0.006  \\
$t_{\rm E}$ (days)             &  130.53  $\pm$ 1.00   &   133.57  $\pm$ 0.53   &   152.10  $\pm$ 0.178   &   150.34  $\pm$ 0.44   \\
$s$                            &    0.845 $\pm$ 0.005  &     0.834 $\pm$ 0.001  &     1.667 $\pm$ 0.001   &     1.646 $\pm$ 0.007  \\
$q$                            &    0.147 $\pm$ 0.003  &     0.149 $\pm$ 0.001  &     0.232 $\pm$ 0.001   &     0.223 $\pm$ 0.004  \\
$\alpha$ (rad)                 &    1.344 $\pm$ 0.004  &    -1.348 $\pm$ 0.004  &    -1.339 $\pm$ 0.002   &     1.341 $\pm$ 0.003  \\
$\rho$ ($10^{-3}$)             &    1.75  $\pm$ 0.01   &     1.72  $\pm$ 0.01   &     1.70  $\pm$ 0.01    &     1.73  $\pm$ 0.01   \\
$\pien$                        &   -0.049 $\pm$ 0.008  &     0.050 $\pm$ 0.008  &    -0.004 $\pm$ 0.003   &     0.013 $\pm$ 0.007  \\
$\piee$                        &    0.038 $\pm$ 0.004  &     0.037 $\pm$ 0.003  &     0.049 $\pm$ 0.002   &     0.038 $\pm$ 0.004  \\
$ds/dt$ (yr$^{-1}$)            &    0.383 $\pm$ 0.013  &     0.372 $\pm$ 0.009  &     0.369 $\pm$ 0.014   &     0.370 $\pm$ 0.044  \\
$d\alpha/dt$ (rad yr$^{-1}$)   &    1.220 $\pm$ 0.021  &    -1.233 $\pm$ 0.003  &    -0.026 $\pm$ 0.002   &     0.006 $\pm$ 0.013  \\
\hline                           
\end{tabular}
\label{table:three}
\tablecomments{ 
${\rm HJD}^\prime={\rm HJD}-2450000$.
}
\end{table*}

\begin{figure}
\includegraphics[width=\columnwidth]{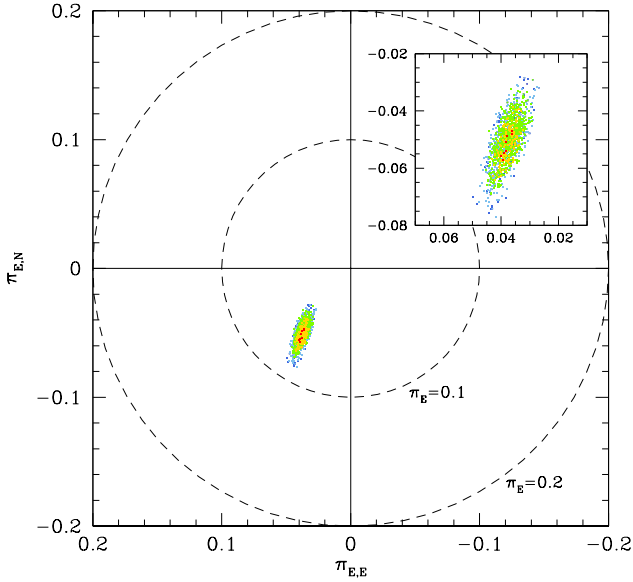}
\caption{
$\Delta\chi^2$ map of MCMC points in the $\pien$ -- $\piee$ parameter plane for the best-fit 
solution (``orbit+parallax'' solution with $u_0>0$).  The color coding represents the points of 
MCMC chain within 1$\sigma$ (red), 2$\sigma$ (yellow), 3$\sigma$ (green), 4$\sigma$ (cyan), and 
5$\sigma$ (blue). 
The inset shows the zoom of the MCMC-point distribution.
}
\label{fig:five}
\end{figure}

We note that the microlens-parallax parameters are well determined despite their small values. 
Figure~\ref{fig:five} shows the distribution of points in the MCMC chain on the $\pien$ -- $\piee$ 
parameter plane. It shows that despite the small magnitude 
$\pie=(\pi_{{\rm E},N}^2+ \pi_{{\rm E},E}^2)^{1/2}\sim 0.06$, the microlens parallax is precisely 
measured to be clearly distinguished from a zero-parallax model. This became possible mainly due 
to the long time scale of the event, which is measured to be $t_{\rm E}\sim 130$~days.

\section{Physical Lens Parameters}\label{section:four}

\subsection{Angular Einstein Radius}\label{section:four-one}

In addition to the microlens parallax, one needs to measure the angular Einstein radius for the 
unique determinations of the lens mass and distance. The angular Einstein radius is measured from 
the normalized source radius by
\begin{equation}
\thetae={\theta_*\over \rho}.
\label{eq3}
\end{equation}
The normalized radius is precisely determined
from modeling. Then, one needs to measure $\theta_*$ to determine $\thetae$.

\begin{figure}
\includegraphics[width=\columnwidth]{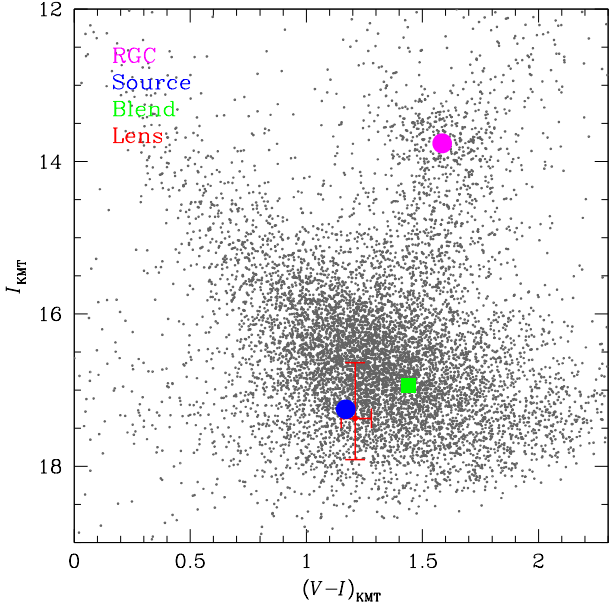}
\caption{
Location of the source and blend with respect to the centroid of the red giant clump (RGC) in 
the instrumental color-magnitude diagram. Also marked is the position of the lens expected from 
the determined mass and distance.
}
\label{fig:six}
\end{figure}

We measure the angular source radius from the de-reddened color $(V-I)_0$ and magnitude $I_0$. In Figure~\ref{fig:six}, 
we mark the position of the source in the color-magnitude diagram (CMD) constructed based on the pyDIA photometry 
of the KMTC data set. To obtain $(V-I)_0$ and $I_0$, we calibrate the instrumental color and magnitude using the 
centroid of red giant clump (RGC), for which its de-reddened color and magnitude $(V-I,I)_{0,{\rm RGC}} = 
(1.06, 14.29)$ \citep{Bensby2011, Nataf2013} are known, as a reference \citep{Yoo2004}. The locations of the source 
and RGC centroid in the instrumental CMD are $(V-I,I)_{\rm S}=(1.17, 17.25)$ and $(V-I,I)_{\rm RGC}=(1.59, 13.76)$, 
respectively.  Then, the de-reddened color and magnitude of the source are estimated from the offsets in color 
$\Delta(V-I)=(V-I)_{\rm S}-(V-I)_{\rm RGC}$ and magnitude $\Delta I = I_{\rm S} - I_{\rm RGC}$ with respect to the 
RGC centroid by $(V-I,I)_{0,{\rm S}}=[(V-I)_{0,{\rm RGC}}+\Delta(V-I), I_{0,{\rm RGC}}+\Delta I] =(0.65\pm 0.02, 
17.78\pm 0.01)$.  The estimated de-reddened color and magnitude indicates that the source is likely 
to be a metal-poor turnoff star.  The measured $V-I$ color is converted into $V-K$ color using the 
color-color relation of \citet{Bessell1988} and the angular source radius is estimated using the relation between 
$V-K$ and the surface brightness of \citet{Kervella2004}. The measured angular source radius is 
$\theta_*=1.04\pm 0.08~\mu$as.

In Table~\ref{table:four}, we list the measured angular Einstein radius. We also present the relative lens-source 
proper motions in the geocentric, $\mu_{\rm geo}$, and heliocentric frames, $\mu_{\rm helio}$. The geocentric 
proper motion vector is determined from the measured angular Einstein radius, event time scale, and microlens 
parallax vector $\pivec_{\rm E}=(\pien,\piee)$ by 
\begin{equation}
\muvec_{\rm geo} = {\thetae\over t_{\rm E}}{\pivec_{\rm E}\over \pie}.
\label{eq4}
\end{equation}
The heliocentric proper motion is computed by
\begin{equation}
\muvec_{\rm helio} = \muvec_{\rm geo} + {\bf v}_{\oplus,\perp}{\pi_{\rm rel}\over {\rm au}},
\label{eq5}
\end{equation}
where ${\bf v}_{\oplus,\perp}$ represents the velocity of the Earth motion projected on the sky at $t_0$ and 
$\pi_{\rm rel}=\pi_{\rm L}-\pi_{\rm S}={\rm au}(D_{\rm L}^{-1} - D_{\rm S}^{-1})$ is the relative lens-source 
parallax \citep{Gould2004, Dong2009}.  The angle $\phi$ represents the orientation angle of $\muvec_{\rm helio}$ 
as measured from the north. It is found that the measured Einstein radius $\thetae\sim 0.6$ mas is close to that 
of a typical lensing event. However, the measured relative lens-source proper motion, 
$\mu_{\rm geo}\sim 1.7~{\rm mas}~{\rm yr}^{-1}$, is substantially slower than a typical value of 
$\sim 5~{\rm mas}~{\rm yr}^{-1}$. This indicates that the long time scale of the event is mainly caused by the 
slow relative lens-source motion. From simulation of Galactic lensing events, \citet{Han2018} pointed out that 
majority of long time-scale events are caused by slow relative lens-source proper motions arising due to the 
chance alignment of the lens and source motions. In this sense, OGLE-2017-BLG-0039 is a typical long time-scale 
event.

\subsection{Mass and Distance}\label{section:four-two}

We estimate the mass and distance to the lens using Equations~(\ref{eq1}) and (\ref{eq2}) and list them in 
Table~\ref{table:five}.  Also listed are the projected separation $a_\perp = s D_{\rm L}\thetae$ and the 
projected kinetic-to-potential energy ratio $({\rm KE}/{\rm PE})_\perp$. The projected kinetic-to-potential 
energy ratio is computed based on the projected binary separation $a_\perp$, lens mass $M=M_1+M_2$, and the 
lensing parameters $s$, $ds/dt$, and $d\alpha/dt$ by \begin{equation}
\left( {{\rm KE}\over {\rm PE}} \right)_\perp =
{ (a_\perp/{\rm au})^3 \over 8\pi^2(M/M_\odot)}
\left[ \left( {1\over s}{ds/dt\over {\rm yr}^{-1}}\right)^2
+
\left( {d\alpha/dt\over {\rm yr}^{-1}} \right)^2\right].
\label{eq6}
\end{equation}
The ratio should be less than unity for the lens system to be a gravitationally bound system. It is found 
that the measured ratios for both the $u_0>0$ and $u_0<0$ solutions satisfy this condition.  Furthermore, 
the estimated kinetic-to-potential energy ratio is within the expected range 
$0.2\lesssim ({\rm KE/PE})_\perp\lesssim 0.5$ for binaries with moderate orbital eccentricity that are not 
viewed at unusual angles, e.g., a wide binary in an edge-on orbit.

\begin{table}
\centering
\caption{Angular Einstein radius and the relative lens-source proper motion}
\begin{tabular}{m{4cm} m{2.5cm} }
\hline\hline
\multicolumn{1}{c}{Parameter} &
\multicolumn{1}{c}{Value} \\
\hline
$\thetae$ (mas)                    &  0.59 $\pm$ 0.04           \\
$\mu_{\rm geo}$ (mas yr$^{-1}$)    &  1.67 $\pm$ 0.12           \\
$\mu_{\rm helio}$ (mas yr$^{-1}$)  &  1.64 $\pm$ 0.12           \\
$\phi$                             &  142$^\circ$ (35$^\circ$)  \\
\hline
\end{tabular}
\label{table:four}
\tablecomments{
The orientation 
angle values of the heliocentric proper motion, $\phi$, in and out of the parenthesis are for the $u_0>0$ and 
$u_0<0$ solutions, respectively. 
}
\end{table}

According to the best-fit solution, the estimated masses of the primary, $M_1\sim 1.0~M_\odot$, 
and the companion, $M_2\sim 0.15~M_\odot$, of the lens correspond to the masses of an early G-type and a late 
M-type dwarfs, respectively. The estimated distance to the lens is $D_{\rm L}\sim 6.0$ kpc.

\subsection{Flux from the Lens}\label{section:four-three}

The estimated mass of the primary lens, $M_1\sim 1.0~M_\odot$, is substantially heavier than 
those of the most common lens population of M dwarfs.  If the lens is a star, then, its flux 
will contribute to the blended flux.

In Figure~\ref{fig:six}, we mark the position of the primary lens, which dominates the flux from 
the binary lens, in the instrumental CMD expected from the determined lens mass and distance.  
Here we assume that the primary lens is a main-sequence star and the ranges of the color and 
magnitude are estimated based on the uncertainty of the estimated lens mass.  Because the estimated 
distance to the lens, $D_{\rm L}\sim 6$~kpc, indicates that the lens is likely to be located behind 
most obscuring dust in the disk, the lens experiences extinction and reddening similar to those of 
the source star.  Under this assumption, we estimate the lens position in the CMD by
\begin{equation}
(V-I,I)_{\rm L}=(V-I,I)_{0,{\rm L}} + [(V-I,I)_{\rm RGC}-(V-I,I)_{0,{\rm RGC}}], 
\label{eq7}
\end{equation}
where $(V-I,I)_{\rm RGC}=(1.59, 13.76)$ represent the apparent color and magnitude of the RGC centroid 
in the instrumental CMD and $(V-I,I)_{0,{\rm RGC}}=(1.06, 14.29)$ represent the de-reddened values. The 
color $(V-I)_{0,{\rm L}}\sim 0.68$ represents the intrinsic color corresponding to the estimated lens 
mass \citep{Allen1976}.  The de-reddened lens magnitude is estimated by
\begin{equation}
I _{0,{\rm L}} =[M_V -(V-I)_{0,{\rm L}}]+5\ \log (D_{\rm L}/{\rm pc}) -5,   
\label{eq8}
\end{equation}
where $M_V\sim 4.7$ is the absolute $V$-band magnitude of a main-sequence star with a mass corresponding 
to the lens mass \citep{Allen1976}. The estimated color and magnitude of the lens in the instrumental 
CMD are $(V-I,I)_{\rm L}=(1.21_{-0.06}^{+0.07}, 17.37_{-0.73}^{+0.54})$. 
Also marked in the CMD is the location of the blend (square dot) which has color and brightness 
of $(V-I,I)_{b}=(1.44, 16.93)$.  It is found that the position of the lens in the CMD is close 
to that of the blend.

\begin{table}
\centering
\caption{Physical lens parameters.}
\begin{tabular}{m{2.5cm} m{2.0cm}  m{2.0cm}}
\hline\hline
\multicolumn{1}{c}{Parameter} &
\multicolumn{1}{c}{$u_0>0$} &
\multicolumn{1}{c}{$u_0<0$} \\
\hline
$M_1$ ($M_\odot$)             &  1.03 $\pm$ 0.15  &  1.03 $\pm$ 0.15 \\
$M_2$ ($M_\odot$)             &  0.15 $\pm$ 0.02  &  0.15 $\pm$ 0.02 \\
$D_{\rm L}$ (kpc)             &  5.99 $\pm$ 0.77  &  5.94 $\pm$ 0.76 \\
$a_\perp$ (au)                &  3.00 $\pm$ 0.39  &  2.99 $\pm$ 0.38 \\
$({\rm KE}/{\rm PE})_\perp$   &  0.49 $\pm$ 0.02  &  0.49 $\pm$ 0.02 \\
\hline
\end{tabular}
\label{table:five}
\end{table}

The proximity of the lens position to that of the blend in the CMD indicates that the flux 
from the lens comprises a significant fraction of the blended flux.  From the comparison 
of the $I$-band magnitudes of the lens and blend, it is found that the lens is 
$\Delta I\sim 0.44$ magnitude fainter than the total blended flux. Because the flux from 
the lens contributes to the blended flux, this means that the lens comprises $\sim 2/3$ of 
the $I$-band blended flux, although this fraction is somewhat uncertain due to the 
uncertainty of the lens brightness.  The lens is bluer than the blend by $\Delta(V-I)\sim 0.23$.  
This indicates the possibility that there may exist another blended star that is redder than the 
lens.  An alternative explanation of the slight differences in color and brightness between the 
lens and blend is that the lens is a ``turnoff star'' rather than a main-sequence star.  Since a 
turnoff star has evolved off the main sequence, it is slightly redder and brighter than a main 
sequence star with a same mass.

In order to check the two possible explanations for the slight differences in the color and 
brightness between the lens and blend, we measure the ``astrometric offset'' between the 
source and baseline object using the OGLE images.  If there exists another blend, the position 
of the baseline object, which corresponds to the centroid of the combined image of the blend and 
source, would be slightly different from the position when the source is magnified, and thus there 
would be an astrometric offset.  If the ``turnoff star'' explanation is correct, on the other hand, 
there would be no measurable astrometric offset.  From this measurement, we find that the image 
centroid when the source was magnified is shifted from the baseline object by $\sim 0.26 \pm 0.07$ 
pixels, which corresponds to $\sim 65\pm 17$ mas.  This supports the explanation that there is another 
blend.

\section{Conclusion}\label{section:five}

We analyzed the binary-lensing event OGLE-2017-BLG-0039.  The long duration of the event enabled 
us to precisely measure the microlens parallax despite the small value.  In addition, the analysis 
of the well resolved caustic crossings during both source star's entrance and exit of the caustic 
allowed us to measure the angular Einstein radius.  From the combination of $\pie$ and $\thetae$, 
we measured the mass and distance to the lens and found that the lens was a binary composed of an 
early G and late M dwarfs located at a distance $\sim 6$ kpc.  From the location of the lens in the 
color-magnitude diagram, it was found that the flux from the lens comprised $\sim 2/3$ of the blended 
light.  Therefore, the event was a rare case of a bright lens event for which follow-up spectroscopic 
observations could confirm the nature of the lens.

\acknowledgements
Work by C.H.\ was supported by the grant (2017R1A4A1015178) of
National Research Foundation of Korea.
The MOA project is supported by JSPS KAKENHI Grant Number JSPS24253004, JSPS26247023,
JSPS23340064, JSPS15H00781, and JP16H06287.
The OGLE project has received funding from the National Science Centre, Poland, grant 
MAESTRO 2014/14/A/ST9/00121 to A.~Udalski.  
Work by A.G. was supported by JPL grant 1500811 and US NSF grant AST-1516842.
Work by J.C.Y.\ was performed under contract with
the California Institute of Technology (Caltech)/Jet Propulsion
Laboratory (JPL) funded by NASA through the Sagan
Fellowship Program executed by the NASA Exoplanet Science Institute.
Work by Y.S.\ was supported by an appointment to the NASA Postdoctoral Program at the Jet
Propulsion Laboratory, California Institute of Technology, administered by Universities Space
Research Association through a contract with NASA.
This research has made use of the KMTNet system operated by the Korea
Astronomy and Space Science Institute (KASI) and the data were obtained at
three host sites of CTIO in Chile, SAAO in South Africa, and SSO in
Australia.
We acknowledge the high-speed internet service (KREONET)
provided by Korea Institute of Science and Technology Information (KISTI).

\end{document}